\documentclass[12pt]{article}
\usepackage{graphicx}

\newcommand{\beq}{\begin{equation}}
\newcommand{\eeq}{\end{equation}}
\newcommand{\beqs}{\begin{eqnarray}}
\newcommand{\eeqs}{\end{eqnarray}}

\newcommand{\Sec}[1]{Sec.~\ref{#1}}
\newcommand{\Eq}[1]{Eq.~(\ref{#1})}

\begin{document}
\baselineskip 0.6cm

\def\simgt{\mathrel{\lower2.5pt\vbox{\lineskip=0pt\baselineskip=0pt
           \hbox{$>$}\hbox{$\sim$}}}}
\def\simlt{\mathrel{\lower2.5pt\vbox{\lineskip=0pt\baselineskip=0pt
           \hbox{$<$}\hbox{$\sim$}}}}

\begin{titlepage}
\begin{flushright}
MIT-CTP/4135
\end{flushright}

\vskip 2.0cm

\begin{center}

{\Large \bf 
Dark matter from dynamical SUSY breaking 
}

\vskip 1.0cm
{\large JiJi Fan$^1$, Jesse Thaler$^2$, and Lian-Tao Wang$^1$}

\vskip 0.4cm

$^1$ {\it Department of Physics, Princeton University, Princeton, NJ 08540} \\
$^2$ {\it Department of Physics, Massachusetts Institute of Technology, Cambridge, MA, 02139} \\

\vskip 2.0cm

\abstract{We consider explicit models of dynamical supersymmetry breaking where dark matter is a 10 -- 100 TeV strongly-interacting composite state carrying no standard model quantum numbers.  These constructions are simple variants of well-known supersymmetry breaking mechanisms, augmented to allow for a large ``flavor'' symmetry.  Dark matter is the lightest composite modulus charged under this symmetry and is a viable cold dark matter candidate with a thermal relic abundance.  This is an attractive possibility in low-scale gauge-mediated scenarios where the gravitino is the lightest superparticle.  A light $R$-axion associated with supersymmetry breaking is present in these hidden sectors and serves as the portal between dark matter and the standard model.  Such scenarios are relevant for present and future indirect detection experiments.
}

\end{center}
\end{titlepage}

\setcounter{tocdepth}{2}

\section{Introduction}

The nature of the dark matter (DM) in the universe is a central open question in particle physics, astrophysics, and cosmology.  A leading paradigm for the origin of DM is thermal freezeout of stable massive particle, which yields an appropriate relic abundance of cold dark matter \cite{Goldberg:1983nd,Ellis:1983ew}.  The most commonly studied case is a weakly-interacting massive particle (WIMP), where DM has a mass of 100 GeV -- 1 TeV and annihilates through weak couplings. However, there is a much larger range of DM masses and couplings consistent with thermal freezeout.

In particular, strongly-coupled DM can have the right relic abundance if its mass is in the range 10 -- 100 TeV.  This can occur if DM is a stable state in a hidden sector which has small direct couplings to the standard model (SM)~\cite{Dimopoulos:1996gy, Nomura:2005, Hamaguchi:2007}.  Indeed, there is independent motivation for this 10 -- 100 TeV mass scale coming from supersymmetry (SUSY), specifically low-scale gauge mediation.  An important ingredient for any model of low-scale mediation is a dynamical SUSY breaking (DSB) sector, and in the lowest-scale mediation models, the DSB scale is $\Lambda \sim 10$ -- 100 TeV.  Here, we study the possibility that DM is a stable state in the DSB sector with no SM quantum numbers, with a mass determined by this dynamical scale.

To set the stage, recall that gauge-mediated supersymmetry breaking (GMSB)~\cite{Dine:1993yw} is an appealing paradigm in its own right since it allows for natural suppression of flavor violation in the supersymmetric standard model (SSM)~\cite{gmsb}.   However, in low-scale GMSB, there is no good candidate for cold DM among the SSM fields.  The lightest superpartner is almost always the gravitino,\footnote{There are notable exceptions, see Refs.~\cite{Nomura:2001ub}.} which can only be a warm DM candidate if its mass $m_{3/2}$ is about a keV.\footnote{In addition, structure formation typically requires the gravitino mass to be lighter than $\mathcal{O}(10~\mathrm{eV})$ \cite{Viel:2005qj}.}  So in the context of GMSB, a stable state in the DSB sector is compelling candidate for cold DM.

Simply having strongly-coupled DM is insufficient for DM to be a thermal relic; there must be sufficiently strong interactions between the DSB and the SSM to maintain thermal equilibrium before freezeout.  As advocated in Ref.~\cite{Mardon:2009gw}, strongly-coupled DM can couple to the SSM via a light mediator field, dubbed a ``portal''.  Since DSB is almost always associated with spontaneous breakdown of an $R$ symmetry~\cite{Nelson:1993nf}, a natural candidate for this portal field is the light $R$-axion.  Moreover, if DM is only quasi-stable with a lifetime $\tau \simeq 10^{26}~\mathrm{sec}$, then DM can decay via this $R$-axion~\cite{Mardon:2009gw}, potentially explaining the electron/positron excesses seen in cosmic ray observations~\cite{Papini:2009zz, Abdo:2010nc, Acero:2009zz, Aharonian:2009ah}.

In this paper, we study DM models based on different DSB mechanisms, and discuss the necessary conditions for the existence of a viable DM candidate.  In addition to having a portal field like the $R$-axion, DM in the DSB sector requires a global symmetry preserved by the strong dynamics, such that DM can be the lightest state charged under this symmetry.  While the simplest DSB models frequently considered in the literature do not satisfy these conditions (e.g.\ the 3-2 or 4-1 models~\cite{Dine:1993yw, Affleck:1984xz, 6-1}), we find there are straightforward generalizations in which composite DSB states can be viable DM candidates.  

Concretely, we consider two explicit models with viable DM based on two popular DSB mechanisms:
\begin{itemize}
\item SUSY breaking from a dynamically generated superpotential~\cite{Dine:1993yw, Affleck:1984xz, 6-1};
\item SUSY breaking from a quantum moduli space (QMS)~\cite{itiy}.
\end{itemize} 
We also briefly mention DM in the context of metastable SUSY breaking~\cite{iss}.  Since we will be considering these DSB sectors in the strongly-coupled limit, we will only be able to study the symmetry structures of these sectors and estimate their DM properties.  However, finding DSB sectors that satisfy the necessary conditions for composite DM is already interesting, and we expect that these requirements will open new directions in DSB sector model building.  

We will show that under reasonable assumptions about the DSB sector, DM can be cosmologically stable and have the desired thermal relic abundance.  While not the main focus of this work, we will also show that such models may be relevant for indirect detection experiments, in particular the possibility of DM decay being the source of several recently observed excesses in PAMELA~\cite{Papini:2009zz}, FERMI~\cite{Abdo:2010nc} and H.E.S.S.~\cite{Acero:2009zz, Aharonian:2009ah} (see Refs.~\cite{Mardon:2009gw,Nardi:2008ix, Cholis:2008vb, Arvanitaki:2008hq,Chen:2009iua}).  This occurs if the symmetries in the DSB sector that stabilize DM are violated by physics at the unification or Planck scale, and our explicit constructions naturally accommodate this possibility.

This paper is organized as follows.  In \Sec{sec:overview}, we discuss the desired properties of DM from the DSB sector and briefly mention alternative scenarios.  We then construct two explicit models in \Sec{sec:32} and \Sec{sec:itiy}, showing that two popular DSB mechanisms allow the necessary symmetry structures to have viable DM.  In \Sec{sec:messengers}, we briefly discuss how such models can fit into the GMSB framework by introducing messenger fields.  We conclude in \Sec{sec:conclude}. 

\section{Viable dark matter from DSB}
\label{sec:overview}

\subsection{Minimal requirements}
\label{sec:minimal}

Our starting point is a DSB sector with fields and interactions characterized by a scale $\Lambda \sim$ (10 -- 100 TeV).  This scale is determined by the strong dynamics that triggers SUSY breaking.  In order for the DSB sector to have a DM candidate, it must possess at least one (accidental) global symmetry which is left unbroken by the strong dynamics.  The lightest state charged under this global symmetry is a stable DM candidate, and the natural size for its mass $m_{\rm DM}$ is the SUSY breaking scale $\Lambda$.  We briefly mention other possibilities in \Sec{sec:altconstruct}.

The relic abundance of such a composite is determined from its annihilation cross section into other lighter strongly-interacting particles not carrying the conserved quantum number, which are in turn kept in thermal equilibrium with SM degrees of freedom through some kind of portal.\footnote{In GMSB models, thermal equilibrium might also be maintained through couplings to the messenger sector.}  Assuming $2 \rightarrow 2$ annihilation, the thermally averaged cross section is 
\begin{equation}
\label{eq: annihilation xsec}
\langle \sigma v \rangle \approx \frac{1}{8 \pi} \frac{\kappa^4}{m_{\rm DM}^2},
\end{equation}
where $\kappa$ denotes the relevant coupling of the hidden sector.  As recently emphasized in Ref.~\cite{Mardon:2009gw}, for typical couplings of a strongly-coupled sector $\kappa \sim (\sqrt{4\pi}-4\pi)$ and $m_{\rm DM}\sim {\cal{O}}(10-100\, \rm{TeV})$, we obtain the desired relic abundance.\footnote{The right relic abundance could also be maintained if the DM mass is lower than the weak scale, and if the relevant couplings is smaller, see Ref.~\cite{Feng:2008ya}.}   With SUSY breaking (and messenger fields) at this scale $\sqrt{F} \sim {\cal{O}}(10-100\, \rm{TeV})$, gauge mediation generates MSSM gaugino and scalar masses of order $(g^2 / 16 \pi^2) \sqrt{F} \approx {\cal{O}}(100\, \rm{GeV} - 1\, \rm{TeV})$.  The gravitino mass is $m_{3/2} = F/ \sqrt{3}M_{\rm Pl} \approx (0.1- 10\, \rm{eV})$, where $M_{\rm Pl} \sim 10^{18}\, \rm{GeV}$ is the reduced Planck scale, and the resulting gravitino abundance is small and consistent with structure formation~\cite{Viel:2005qj}. 

A simple way to keep the DSB sector in thermal equilibrium with the SSM is to assume that both sectors are charged under the same (approximate) global symmetry which is subsequently broken spontaneously \cite{Mardon:2009gw}.  In SUSY breaking models, $R$ symmetry is a natural candidate, and we will focus on that case throughout.  Generic DSB models will spontaneously break this $R$ symmetry~\cite{Nelson:1993nf}, and as long as there are no large sources of explicit $R$ violation, the resulting $R$-axion will be light.  The $R$-axion is then coupled to both the DSB and SSM sectors, and will keep the two sectors in thermal equilibrium.  Intriguingly, supergravity yields an irreducible contribution to the $R$-axion mass, since to cancel the cosmological constant, the superpotential must have a $R$ symmetry breaking constant term $FM_{\rm Pl}$~\cite{Bagger:1994hh}.  After taking into account of possible additional explicit $R$ symmetry breaking effects, the $R$-axion mass is then in a phenomenologically interesting range $m_a \approx  {\cal{O}}(1\, \rm{MeV} - 10\, \rm{GeV})$, with implications for collider experiments and dark matter indirect detection \cite{Mardon:2009gw}.  

To summarize, the minimal symmetry requirements to have strongly-coupled thermal relic DM in the DSB sector are:
\begin{enumerate}
\item An (approximate) unbroken global symmetry under which DM is charged.
\item A spontaneously broken (approximate) global symmetry under which both the SSM and DSB sectors are charged.  The natural candidate is an $R$ symmetry, resulting in a $R$-axion.  
\end{enumerate}
Note that this second requirement of an ``axion portal'' \cite{Nomura:2008ru} is not unique.  For example, one could maintain thermal equilibrium between the DSM and SSM sectors using the vector $U(1)$ portal~\cite{ArkaniHamed:2008qp}, which we will briefly mention in \Sec{sec:altconstruct}.

\subsection{Strong coupling}

For the large values of $\kappa$ necessary to get the right relic abundance, one might worry about maintaining theoretical control over the DSB dynamics.  In general, one is not able to calculate the K$\ddot{\rm{a}}$hler potential in a strongly-coupled SUSY theory.  Thus, the details of the low-energy theory below the strong dynamics scale---such as the masses of low-lying states and the proper identification of unbroken global symmetries---are strictly speaking unknown.  Our strategy is to identify DSB models which satisfy the necessary conditions to allow a DM candidate, and assume that the symmetry structure at small values of $\kappa$ is preserved at strong coupling.  

Indeed, it is already non-trivial to identify explicit weakly-coupled DSB models that possess the necessary symmetries to have stable DM.   While the above requirements in \Sec{sec:minimal} were already outlined in Ref.~\cite{Mardon:2009gw}, that work could not identify an explicit single-scale DSB model with viable DM.  In particular, Ref.~\cite{Mardon:2009gw} was only able to show that a dynamical (but SUSY-preserving) sector could be consistently coupled to an $R$-breaking and SUSY-breaking O'Raifeartaigh model.  In the present work, we will show two examples of weakly-coupling DSB sectors with the right structure to contain DM, and then extrapolate to strong coupling.

A key assumption we will make in our studies is that the relevant degrees of freedom for describing DM are gauge-invariant moduli fields.  However, since we are considering the limit where $m_{\rm DM} \simeq \Lambda$, in principle other composite degrees of freedom might be the true DM modes.  That said, since there is always a parametric limit where the moduli can be made lighter than the other composites, we think this assumption is justified.

\subsection{Dark matter detection}
\label{subsec:generalDMdetection}

Though not the main focus of this paper, we will show that DSB DM models may be relevant for DM indirect detection experiments.  In particular, recent cosmic ray excesses~\cite{Papini:2009zz, Abdo:2010nc, Acero:2009zz, Aharonian:2009ah} might be explained by heavy DM decay to the $R$-axion~\cite{Mardon:2009gw}.  This can occur if the global symmetries protecting DM are only accidental at low energies, and are not respected by physics at high scales $M_*$ such as the unification or Planck scale~\cite{Arvanitaki:2008hq}.  In the low energy effective theory, these breakings are encoded in high dimension operators suppressed by $M_*$, which can induce DM decay, and we will see that DSB models naturally accommodate such operators. We emphasize that it is possible that such operators are absent in the DSB DM  models considered here. In this case,  the stable state in the model still provides viable DM candidate. Although it is not possible to explain the recent cosmic ray excesses, the annihilation of dark matter can still give indirect detection signals which may be probed by future observations. 

The DM lifetime which can potentially explain recent cosmic ray excesses is of order $10^{26}$ sec~\cite{Arvanitaki:2008hq}.   For 1 -- 100 TeV DM, dimension-six symmetry breaking operators suppressed by the unification or Planck scale could explain the anomalous astrophysical signals~\cite{Arvanitaki:2008hq}.  For $R$-axion decay constants in the range $f_a \approx {\cal{O}}(1-100\, \rm{TeV})$, the most natural region avoiding all the constraints from astrophysics and rare meson decays are $m_a>2m_\mu$~\cite{Mardon:2009gw}.   For the range $m_a \approx (200\, \rm{MeV} - 10\, \rm{GeV})$, the $R$-axion decays mostly into either $\mu^+\mu^-,\tau^+\tau^-$, or $\pi^+\pi^-\pi^0$ and provide a good fit for the electron/position signals in cosmic rays~\cite{Mardon:2009gw}.

In the scenario considered in this paper, it is difficult for direct detection experiments to discover DM if it is not charged under SM gauge groups.  The SM-singlet DM can only interact with nuclear matter via $R$-axions or loop diagrams mediated by messenger fields, and the resulting cross sections are unobservable with present techniques.

\subsection{Alternative constructions}
\label{sec:altconstruct}

In this paper, we will focus on the possibility that DM is a 10 -- 100 TeV state in the DSB sector with no SM quantum numbers.  Here, we briefly mention some alternative possibilities for DM in GMSB-like scenarios.

First, the messenger sector---either a separate sector or part of the DSB sector---has previously been proposed as a possible source for a cold DM candidate in GMSB models~\cite{Dimopoulos:1996gy, Hamaguchi:2007, messenger dm, Hamaguchi:2008rv}.  The lightest messenger particle could be stable provided the existence of a ``messenger number" invariance.  Assuming that they only annihilate into particles of the SSM through SM gauge interactions, they should be lighter than about 5 TeV in order not to overclose the Universe.  Direct detection experiments have already ruled out such a WIMP~\cite{Ahmed:2009zw} unless the elastic scattering of DM off nucleons is forbidden~\cite{messenger dm}.  Another possibility is to have a strongly-coupled composite heavy messenger at the scale of 10 -- 100 TeV as DM~\cite{Hamaguchi:2008rv}.  As we mention in \Sec{sec:messengers}, even if DM dominantly arises from the DSB sector, the messenger sector often has stable states of its own, which must be taken into account.

In our constructions, we focus on the DSB sector, and we will be considering the case that the mass of DM mass is comparable to the dynamical scale.  However, it is also possible for the DSB sector to contain stable states which are parametrically lighter than the DSB scale, e.g., pseudo-moduli or pseudo Nambu-Goldstone bosons (pNGB).  They can potentially be good cold DM candidates, as has been recently explored in the literature~\cite{Ibe:2009dx,Shih:2009he,KerenZur:2009cv}.  From the basic requirements for DM such as the correct relic abundance, though, it seems necessary to extend beyond the basic ingredients of the DSB sector in order for such scenarios to be viable.

For example, a weakly-coupled Intriligator-Thomas-Izawa-Yanagida (ITIY) model with scalar mesons as DM has been studied in~\cite{Ibe:2009dx}. The scalar mesons are pNGBs with suppressed couplings $\lambda \sim {\cal{O}}(10^{-2} -10^{-4})$ and masses around TeV.  To achieve the right amount of relic abundance, a ``resonance mechanism" was invoked by tuning the mass of the pseudo-modulus $S$ to be very close to twice the meson DM mass.  Once the proper relic abundance is achieved, possible DM decays in Ref.~\cite{Ibe:2009dx} were induced by a holomorphic K\"{a}hler potential terms $K \sim \Lambda  M_i$ or equivalently dimension five superpotential terms after taking into account of supergravity, similar to what we will do in \Sec{subsec:DMdecayITIY}.

Both pseudo-moduli and pNGBs could be DM candidates in models with metastable SUSY breaking~\cite{Shih:2009he,KerenZur:2009cv}.  It has recently been shown by Intriligator, Seiberg and Shih (ISS) that deformed SQCD theories with weakly-coupled magnetic duals have a meta-stable SUSY breaking vacuum near the origin of field space~\cite{iss}.  The original ISS model has massless fermionic moduli fields and some pNGBs from flavor symmetry breaking (baryons or dual quarks).  Both are charged under flavor symmetries (baryons are further charged under a $U(1)_B$) and the lightest states are stable. The annihilation of the moduli to pNGBs or vice versa will be suppressed by couplings much smaller than ${\cal{O}}$(1), and the resulting relic density tends to overclose the universe.  Thus to have a viable DM in ISS models, one has to make the fermionic moduli massive and have additional interactions for DM to annihilate more efficiently. 

We will not try to build a detailed DM model based on the ISS model but just point out possible solutions.  First, to make the fermionic modulus massive, one must break an accidental $R$ symmetry in the ISS model.  This can be understood in analogy to the gaugino masses in the gauge mediation, since both fermion mass terms break a continuous $R$ symmetry.  Thus both fermions can acquire masses at one loop after $R$ symmetry is broken.  $R$ symmetry can be broken either explicitly by additional fields and operators or spontaneously through the ``inverted hierarchy" mechanism~\cite{r breaking}. 

Second, to have large enough DM annihilation cross section in the ISS model, one could gauge some unbroken flavor symmetries. For instance, in the $N_f=N_c+1$ ISS model, we can adjust the parameters such that the baryon is the lightest field charged under the global symmetries. Then, we can gauge the $U(1)_B$ which can kinematically mix with the SM hypercharge $U(1)_Y$.   The DM sector then communicates to the SM through the $U(1)$ vector portal~\cite{ArkaniHamed:2008qp, Shih:2009he}, and it is possible to have the right DM relic density.  Another model is constructed in Ref.~\cite{iss dm} where the unbroken flavor symmetry is gauged and identified with the SM gauge group.  The fermionic pseudo-moduli then annihilates into the SM fields through SM gauge interactions.

\section{Dark matter from a dynamical superpotential}
\label{sec:32}

In this section, we show that the requirements from \Sec{sec:minimal} to achieve DM in the DSB sector can be met in models with a non-perturbatively generated superpotential.   Many early models of DSB were based on such constructions, e.g., the 3-2 model~\cite{Affleck:1984xz} and the 4-1 model~\cite{Dine:1993yw, 6-1}.  In these models, a non-perturbative superpotential lifts the origin of field space while an added classical superpotential constrains some fields to stay near the origin.  SUSY breaking can be established by the Affleck-Dine-Seiberg (ADS) criterion~\cite{Affleck:1984xz, Affleck:1983vc}:  the classical scalar potential lifts all non-compact flat directions and there exists a spontaneously broken global symmetry (an $R$ symmetry in this case).

The existence of a spontaneously broken $R$ symmetry meets one of the criteria for DM in the DSB sector.  However, the 3-2 and 4-1 models do not have additional unbroken global symmetries that would lead to stable DM.  As we will see, by extending the symmetry structure of these models one can get viable composite DM.  Our construction requires a large unbroken global symmetry, so in this sense, it is similar in spirit to ``single sector'' models where SSM degrees of freedom are composites from a DSB sector~\cite{ArkaniHamed:1997fq}.

\subsection{General considerations}

Recall that the 3-2 and 4-1 models are the simplest examples of two infinite classes of DSB models based on $SU(n) \times SU(2) \times U(1)$ ($n$ odd) and $SU(n) \times U(1)$ ($n$ even) gauge groups.  They can be constructed using the discarded generator method~\cite{Dine:1993yw, Leigh:1997sj}.  For example, $SU(n) \times U(1)$ ($n$ even) models can be deduced from the non-calculable theory with gauge group $SU(n+1)$, an antisymmetric tensor, and $n-3$ antifundamentals.  After discarding those generators of $SU(n+1)$ which do not lie in the subgroup $SU(n) \times U(1)$, the original chiral fields decompose as 
\begin{equation}
A_2+F_{1-n}+(n-3)\bar{F}_{-1}+(n-3)S_n,
\end{equation}
where $A$ is an antisymmetric tensor, $F(\bar{F})$ is the (anti)fundamental, $S$ is a singlet of $SU(n)$, and the subscripts denote the $U(1)$ charges.

When $n \ge 6$, these models have an $SU(n-3)$ global symmetry under which $\bar{F}$ and $S$ are charged.  At the classical level, one can add to this model a superpotential which lifts all flat directions
\begin{equation}
W = \lambda_{ij}A\bar{F}^i\bar{F}^j+\eta_{ij}F\bar{F}^iS^j.
\end{equation}
For suitable choices of the $\lambda$ and $\eta$ matrices, the global $SU(n-3)$ (or some subgroup) will be preserved.  If the symmetry is not broken spontaneously after SUSY is broken, this would signal the existence of stable degrees of freedom to be DM candidates.  The same argument can be applied to the other models constructed in this way.

The minimal models, i.e.\ 3-2 and 4-1, possess no stable degrees of freedom as there is no global symmetry left after SUSY breaking. 
Thus, in order to incorporate a DM candidate, the field content needs to be enlarged to allow for a larger global symmetry which remains unbroken after SUSY breaking.

\subsection{An explicit $SU(6) \times U(1)$ model}
\label{subsec:32explicit}

The simplest example in $SU(n) \times U(1)$ ($n$ even) with a non-Abelian global symmetry is the $SU(6) \times U(1)$ model, and this model does have the right symmetry structure to have stable DM.  The microscopic field content of the $SU(6) \times U(1)$ model is:
\begin{center}
\begin{tabular}{c|cc||cc}
  & $SU(6)$ & $U(1)$ &$SU(3)$& $U(1)_R$ \\
 \hline
  $A^{\alpha\beta}$& $\mathbf{15}$ & 2& $\mathbf{1}$ & $-4$ \\
  $F^{\alpha}$ & $\mathbf{6}$ & $-5$& $\mathbf{1}$ & 3 \\
  $\bar{F}_{\alpha}^i$ & $\mathbf{\bar{6}}$& $-1$&$\mathbf{3}$ & 3 \\
  $S_i$&$\mathbf{1}$ &6&$\mathbf{\bar{3}}$&$-4$ \\
\end{tabular}\,,
\end{center}
where the Greek scripts are the gauge indices and the Roman scripts $i$ $( = 1,2,3)$ are the global $SU(3)$ indices.
At the classical level, one can add a superpotential:
\begin{equation}
\label{eq:classicalMicro61}
W_{cl}=\lambda \epsilon_{123}A^{\alpha\beta}\bar{F}^1_\alpha\bar{F}^2_\beta+\eta_1 F^\alpha(\bar{F}^1_\alpha S_1+\bar{F}^2_\alpha S_2)+\eta_3 F^\alpha \bar{F}^3_\alpha S_3,
\end{equation}
which lifts all the $D$-flat directions as we will show later. This classical superpotential explicitly breaks the global $SU(3)$ to $SU(2)$. Notice that although we choose to preserve $SU(2)$, a global $U(1)^2$ or $U(1)$ would be sufficient to satisfy the minimal requirement for DM inside the DSB sector.

For a range of parameters of the model $\lambda, \eta \ll 1$, the fields obtain vacuum expectation values (VEVs) much above the dynamical scale $\Lambda$ of the $SU(6)$ theory.  The theory is then weakly-coupled and one could analyze SUSY breaking in terms of the microscopic variables since the K$\ddot{\rm{a}}$hler potential of the light degrees of freedom is nearly canonical.  Instead, we focus on the strongly-coupled regime where the scale of SUSY breaking is around the dynamical scale $\Lambda$.  In the effective theory below the scale $\Lambda$, the relevant degrees of freedom are the gauge invariants, i.e.\ moduli, built out of the microscopic fields. Note that $U(1)$ is infrared free and its gauge coupling is weak at the scale $\Lambda$.

The $SU(6)$ invariant moduli fields are 
\begin{equation}
\label{eq:su6moduli}
S_i, \quad M^i= F^{\alpha}\bar{F}_{\alpha}^i, \quad H_i= \epsilon_{ijk}A^{\alpha\beta}\bar{F}_\alpha^j\bar{F}_\beta^k, \quad B \equiv \rm{Pf} \, A=\epsilon_{\alpha_1\alpha_2\alpha_3\alpha_4\alpha_5\alpha_6}A^{\alpha_1\alpha_2}A^{\alpha_3\alpha_4}A^{\alpha_5\alpha_6}. 
\end{equation}
They have charges $S(6)$, $M(-6)$, $H(0)$, $B(6)$ under the gauged $U(1)$.  In general, one expects non-perturbative $SU(6)$ dynamics to cause $M$ or $B$ or both to gets VEVs, breaking the $U(1)$ gauge symmetry. 
The relevant degrees of freedom in the low energy theory are the $U(1)$ invariants built out of the above $SU(6)$ moduli fields: 
\begin{equation}
X^i_j=S_jM^i, \quad H_i, \quad Y^i=M^iB.
\label{moduli}
\end{equation} 
They are not all independent and satisfy the classical constraints:
\begin{equation}
X^i_jY^k-X^k_jY^i=0.
\label{constraint}
\end{equation} 
It can be shown explicitly by global symmetries that the constraint is not modified quantum mechanically.\footnote{That is, $XY$ has $R$-charge $-4$ so it is inconsistent to have a zero $R$-charge instanton term $\Lambda^2$ in the constraint.}  Notice that the 9 constraints in Eq.~(\ref{constraint}) are not all independent. For each fixed $j$, only two of them are independent.  In total, there are 6 independent constraints which allow us to eliminate 6 degrees of freedom from the 15 moduli fields listed in Eq.~(\ref{moduli}). We choose the independent 9 degrees of freedom to be 
\beq
X^3_j : (\mathbf{\bar{2}}+\mathbf{1})_2, \quad H_i : (\mathbf{\bar{2}}+\mathbf{1})_2 , \quad Y^i :(\mathbf{2}+\mathbf{1})_{-6},
\eeq
where we have listed the quantum numbers of the moduli under the unbroken global $SU(2) \times U(1)_R$. Note that we have 10 independent $SU(6)$ moduli in Eq.~(\ref{eq:su6moduli}) before gauging $U(1)$.  After the gauged $U(1)$ eats a multiplet, there should be only 9 independent fields left, which is consistent with our explicit construction and counting above.

In terms of the gauge invariants, we express the classical superpotential as 
\beqs
W_{cl}&=&\lambda H_3+\eta_1  (X^1_1+X_2^2)+\eta_3 X_3^3 \nonumber \\
            &=& \lambda H_3+\eta_1 \left(\frac {X^3_1 Y^1+X_2^3 Y^2}{Y^3}\right)+\eta_3 X_3^3 ,
\label{eq:wcl}
\eeqs
where in the second line, we have expressed everything in terms of the independent moduli fields.  Now consider the microscopic equation of motion (i.e. in terms of $A$, $F$, $\bar{F}$, and $S$) corresponding to this classical superpotential. The equations $\partial W/\partial S$ sets $X,Y$ to zero if we multiply by $S,B$. Similarly $\partial W/\partial \bar{F}$ sets $H$ to zero if we multiply by $\bar{F}$. Thus at the classical level, our theory has no flat directions, satisfying one of the ADS criteria. 

The dynamically generated superpotential is~\cite{6-1}
\begin{equation}
W_{dyn}= \frac{{\Lambda}^{7}}{\sqrt{YH}}.
\label{eq:wdyn}
\end{equation}
The exact superpotential is then a combination of Eqs.~(\ref{eq:wcl}) and (\ref{eq:wdyn}). 
The equation of motion for $X^3_3$ cannot be satisfied, and thus SUSY is broken. In the regime near the origin of the moduli space, the K\"{a}hler potential is smooth in terms of the composite fields, e.g., $K_{\rm eff} \sim |Y|^2 + |H|^2$. Minimizing the potential shows that there is a runaway direction with $Y^3 \to \infty$. But if $\langle Y \rangle$ is large compared to $\Lambda$, we should treat the quarks as elementary degrees of freedom and the K\"{a}hler potential is smooth in $A, F, \bar{F}$. It can be shown that the potential rises when $\langle A, F, \bar{F} \rangle \gg \Lambda$~\cite{Dine:1993yw}. We see that there is no supersymmetric vacuum for either large or small values on the moduli space, and so there could be at least a local SUSY breaking minimum for $\langle Y, H \rangle \sim \Lambda$ where the $U(1)_R$ symmetry is also broken.

Because the theory is strongly coupled, the K\"{a}hler potential for the moduli is in general unknown.  So while we have established that SUSY and $U(1)_R$ must be broken, we cannot say for certain what the fields VEVs are.  We will assume that there is a (local) minimum at the branch where the global $SU(2)$ remains unbroken, which means:
\begin{equation}
\label{eq:61vevs}
\langle H_3, X^3_3, Y^3 \rangle \ne 0, \quad \rm{other\; VEVs}=0.
\end{equation}
This is a reasonable assumption, as points of maximal symmetry are generically the stationary points of the energy.  But one should bear in mind that this could be modified due to our lack of knowledge about strongly-coupled theories.  

\subsection{Spectrum and annihilation}

The stable DM candidate is the lightest component(s) of $X^3_i, Y^i, H_i$ with $i=1,2$. Since we are in a regime where the theory is strongly coupled, it is not possible to directly calculate the annihilation cross section for the DM candidate.  However, we will show that the annihilation cross section is generically of the right size to get the desired relic abundance.

First note that the fields that get VEVs in \Eq{eq:61vevs} are indeed unstable.  The $R$ symmetry is broken spontaneously, thus the model has a light $R$-axion with the decay constant $f_a$ of order $\Lambda$.  The couplings of the modulus to the $R$-axion are fixed by symmetries and can be read off from the modulus kinetic terms~\cite{Mardon:2009gw}.  For the scalar components, the kinetic term is ${\cal{L}}= |\partial_\mu \Phi^\prime + i r_\Phi (\langle \Phi \rangle + \Phi^\prime) (\partial_\mu a)/f_a|^2$, where $\Phi^\prime \equiv \Phi -\langle \Phi \rangle$ in which $\Phi$ represents a the scalar field with non-zero VEV, such as $H_3$, $X^3_3$, and $Y^3$.  $r_\Phi$ is an order one coefficient depending on the $R$-charge of that field. Assuming that all the VEVs are real, this yields a coupling of Re($\Phi)$ to two $R$-axions, and coupling between Re($\Phi$), Im($\Phi$) and an $R$-axion. Therefore, Re($\Phi)$ decays promptly into two $R$-axions, while Im($\Phi)$ decays to three R-axions via an off-shell Re($\Phi)$. SUSY breaking leads to an coupling of the fermionic component of $\Phi$ to its scalar component and a gravitino. This allows the fermions to decay into a gravitino and an $R$-axion. 

The massive $U(1)$ vector multiplet is also unstable. After being higgsed, the $U(1)$ gauge boson and gaugino will obtain a mass of order $g_1\langle \Phi \rangle \sim g_1 \Lambda/(4\pi)$ where $g_1$ is the $U(1)$ coupling. As $g_1 \ll 4\pi$, they are lighter than most of the composites with mass around $\Lambda$ except for the $R$-axion. After the $U(1)$ is broken, gauge-variant operators like $K = g_1V |\Phi|^2$ can be generated with $V$ being the vector multiplet. Expressing the operator in components, one can see that it causes the gauge boson to decay to one $R$-axion and the off-shell unstable scalar $\Phi$, which will decay further to $R$-axions. The gaugino will decay to the off-shell fermionic and scalar $\Phi$, both of which will cascade to $R$-axions.

Contributions to the DM spectrum and annihilation cross section come from two sources: the superpotential and the K\"{a}hler potential. A detailed spectrum is impossible to obtain but we can still estimate the sizes of the masses and interaction strength. After canonically normalizing the composite fields, the superpotential becomes 
\beq
W=\tilde{\lambda} \Lambda^2 H_3+\tilde{\eta}_1 \Lambda^2 \left(\frac {X^3_1 Y^1+X_2^3 Y^2}{Y^3}\right)+\tilde{\eta}_2 \Lambda^2 X_3^3+\frac{\alpha \Lambda^4}{\sqrt{YH}},
\eeq
where the coefficients $\tilde{\lambda},  \tilde{\eta}, \alpha$ absorb unknown factors from normalization. Using naive dimensional analysis (NDA)~\cite{nda}, the sizes of the coefficients in the superpotential and the field VEVs are
\begin{eqnarray}
(\tilde{\lambda}, \tilde{\eta})/(\lambda, \eta) &\sim& g, \quad \quad \quad \quad \quad \quad \quad \alpha \sim g^{3},   \nonumber \\
\langle F_{H,X,Y} \rangle &\sim& g\Lambda^2,  \quad \langle H_3, X^3_3, Y^3 \rangle \sim g\Lambda ,
\end{eqnarray}
where $g \sim (4 \pi)^{-1}$. We will take ($\lambda, \eta$) to be of order one and thus ($\tilde{\lambda}, \tilde{\eta}$) of order $g \sim (4 \pi)^{-1}$.
Expanding around the minimum of the potential, the superpotential gives mass terms for the composites as
\begin{equation}
W \approx \Lambda (Y^iH_i + Y^3H_3 + X^3_i Y^i), \quad i=1,2,
\end{equation}
where we neglected order one coefficients. These leads to supersymmetric masses of order $\Lambda \sim (10 - 100 \,\rm{TeV})$ for nearly all the fields except for $X^3_3$ and a linear combination of $X^3_i$ and $H_i$. But these fields will receive higher order K\"{a}hler corrections we will discuss below.
The interaction terms from the superpotential are
\begin{equation}
W \approx g^{-1} (Y^i H_i (Y^3+H_3) + X^3_iY^i Y^3), \quad i=1,2,
\end{equation}
where again we neglected order one coefficients and the interaction strength is of order $\sim 4\pi$.

As the theory is strongly-coupled, control over the K$\ddot{\rm{a}}$hler potential is lost.  In general, the composites will feel ${\cal{O}}$(1) SUSY breaking effects depending on the details of the K$\ddot{\rm{a}}$hler potential.  For instance, there could be higher-order K$\ddot{\rm{a}}$hler terms compatible with all the unbroken symmetries as 
\begin{equation}
|Y|^4, \quad |X|^4, \quad |Y|^2|H|^2,
\end{equation}
which are suppressed by the QCD scale $\Lambda$. These operators will give non-supersymmetric masses to all composites of order $\Lambda$ and interaction between them with couplings of order $4 \pi$. For instance, $|X^3_3|^4$ will give masses to the scalar component of $X^3_3$ and $(X^3_i X^{3\dagger}_3)^2$ will lead to Majorana mass terms for $X^3_i$.

DM could annihilate into the light vector multiplets. The annihilation cross section is suppressed by the weak couplings $g_1^4$ except for the annihilation into longitudinal modes of the gauge multiplet $V$. In that case, the gauge coupling suppression is compensated by the polarization vector product $(\Lambda^2/m_V^2)^2 \sim (4\pi)^4/g_1^4$ where we take $m_V \sim g_1 \Lambda/(4\pi)$.  In addition, if the unstable fields (e.g.\ $Y^3$) are lighter than the DM particles (e.g.\ $Y^i$), the annihilation will proceed through diagrams like $Y^i Y^i \to Y^3 Y^3$.  With strong couplings from NDA estimates, one expects to obtain the correct relic abundance from such processes.  Notice that we have not included multiplicity factors such as the color factor $N$ in our NDA analysis, which will in general decrease the couplings.\footnote{Since it is the loop expansion factor $\kappa^2N$ that saturates the unitary bound $(4\pi)^2$, the coupling $\kappa$ in Eq.~(\ref{eq: annihilation xsec}) is suppressed by $\sqrt{N}$.}  Thus we do not prefer to work at very large $N$.  On the other hand, the thermal freezeout calculation will be affected by various co-annihilation channels if there exist states of comparable masses~\cite{Griest:1990kh}, which may partially compensate for large $N$ suppressions.  Also, even if the unstable particles are heavier than the stable DM particles, thermal freezeout can still proceed, as long as the mass splittings are not too large~\cite{Griest:1990kh}.

\subsection{Dark matter decays}
\label{sec:32decays}

An intriguing feature of these heavy composite DM models is that the global flavor symmetry is generically only an accidental symmetry at low energies. In particular, higher dimension operators---presumably encoding physics at the unification or Plank scale---can violate the flavor symmetries (and the $U(1)_R$ symmetry) while preserving the gauge interactions. This allows for the possibility of DM decays, but on cosmological timescales.  In our scenario, the decay occurs through the light $R$-axion as in Ref.~\cite{Mardon:2009gw}, so the final states from $R$-axion decays can be mainly leptons, explaining the electron/positron excess observed in PAMELA, H.E.S.S., and FERMI.  The required lifetime of order $10^{26}$ sec is obtained if the decay is caused by a dimension six operator suppressed by the unification or Planck scale. If these dimension six operators break $R$ symmetry, they will also contribute to the $R$-axion mass. However, as the contribution is tiny compared to the irreducible supergravity effect, the $R$-axion remains light.

The global $SU(3)$ symmetry was explicitly broken to $SU(2)$ by the classical superpotential in \Eq{eq:classicalMicro61}.  Assuming this $SU(2)$ is preserved by the strong dynamics, the leading $SU(2)$-violating gauge-invariant operators in this model are:
\begin{equation}
W=\frac{1}{M^2_*} F\bar{F}^i \,{\rm{Pf}}\,A , \quad K=\frac{1}{M_*}(\epsilon_{ij3}A\bar{F}^j\bar{F}^3+ S_i F \bar{F}^3),
\end{equation}
where $M_*$ is the high scale. We show the flavor indices explicitly and take all coefficients to be of order one. Below the scale $\Lambda$, these operators are matched into 
\begin{equation}
W=\frac{\Lambda^4}{M^2_*} c_i Y^i, \quad K=\frac{\Lambda^2}{M_*}c_i(H_i+X^3_i),
\end{equation}
where $i=1,2$. The holomorphic K\"{a}hler terms can be translated into superpotential terms as $W=c_i(\Lambda^4/M_*M_{Pl}) (H^i+X^3_i)$. After including these two terms, the VEVs of the DM fields will be shifted away from zero to:
\begin{equation}
 \langle H_i, Y^i, X^3_i\rangle = c_i \frac{ \Lambda^3}{M_*^2},
 \label{eq:vev}
\end{equation}
where the $c_i$ are coefficients of order ${\cal{O}}(1/4\pi)$ by NDA. 

The tiny VEVs of the composites $H,Y, X$ will induce them to decay into the $R$-axion. If the DM is the scalar component of the composite, we have $H,Y, X \to aa$.  For fermionic DM, one $a$ will be replaced by a gravitino. For the scalar DM decay, the lifetime was estimated as \cite{Mardon:2009gw}
\begin{equation}
\tau_{\rm DM}=8\pi \frac{f_a^4}{m_{\rm DM}^3 v^2}\approx 10^{27}~\mathrm{sec} \left(\frac{1/4\pi}{c}\right)^2\left(\frac{M_*}{10^{18}~\rm{GeV}}\right)^4\left(\frac{10~\rm{TeV}}{\Lambda}\right)^5
\end{equation} 
where $v$ is a typical field VEV as in Eq.~(\ref{eq:vev}) with $c$ a general coefficient of size $c_i$. We have also used $m_{\rm DM} \approx 4 \pi f_a \approx \Lambda$ for the final parametrization.

Thus, we have achieved an explicit DSB model where DM is stabilized by an approximate symmetry of the DSB sector.  DM annihilates through unstable states which decay to $R$-axions.  In addition, DM itself can decay on cosmological scales through small symmetry breaking operators.  In \Sec{sec:messengers}, we explain how this DSB sector can be coupled to a messenger sector to have a complete model of gauge mediation.

\section{Dark matter from a quantum moduli space}
\label{sec:itiy}

In addition to the chiral models discussed in \Sec{sec:32}, there is also a class of vector-like DSB models based on theories with a QMS, e.g.\ the ITIY model~\cite{itiy}.  In these models, SUSY QCD with a QMS is coupled to some gauge singlet fields $S$ with a tree-level superpotential $SM$, where $M$ represents some composite fields from the SQCD sector.  The equations of motion with respect to the singlets force the SQCD degrees of freedom to be at origin, which is not on the QMS.  Thus SUSY is broken. These models usually accommodate large flavor symmetries and some stable fields, which satisfy one of the minimum requirements of DM inside DSB models.  The main complication in ITIY-like models is to establish the spontaneous breaking of an $R$ symmetry, which is essential for both the generation of the MSSM gaugino masses and for DM communicating to the SSM through the $R$-axion portal.

Below we will study the simplest model of this class, the ITIY model.  Such models have previously been studied in the context of DM in Ref.~\cite{Ibe:2009dx}, albeit in the weakly coupled regime.  Here we will focus on the strongly coupled regime.  

\subsection{The ITIY model}
The ITIY model contains an $SU(2)$ gauge symmetry with four fundamentals $Q$, as well as six singlet fields $S_a$.  The fundamentals enjoy an $SU(4) \equiv SO(6)$ global symmetry.  The field content, both in terms of microscopic fields and $SU(2)$ invariant ``mesons'' is
\begin{center}
\begin{tabular}{c|c||cc}
  & $SU(2)$ &$SU(4)\equiv SO(6)$& $U(1)_R$ \\
 \hline
  $Q$& $\mathbf{2}$ & $\mathbf{4}$& 0 \\
  $S$ & $\mathbf{1}$ & $\mathbf{6}$  & 2\\
 \hline
  \hline
 $M_a=Q^2$& $\mathbf{1}$ & $\mathbf{6}$ &0 \\
 $S_a$& $\mathbf{1}$ & $\mathbf{6}$&2 
\end{tabular}\,,
\end{center}
where $a=0,1,2,...5$, and we have indicated the non-anomalous $R$ symmetry.  To this model, we add a classical superpotential of the form 
$W_{cl}=\lambda^{ij}S_iM_j$. To satisfy the quantum modified constraint, at least one of the $M_a$ will have to acquire a VEV, which spontaneously breaks the global symmetry $SO(6) \rightarrow SO(5)$. In order to remove the resulting  massless Goldstone mode, the superpotential must contain explicit global symmetry breaking. The minimal choice, which we will make for our model, is
\begin{equation}
W_{cl}=\lambda_0  S_0M_0+\lambda_1 S_iM_i, \quad i=1,...5.
\end{equation}
It explicitly breaks the global $SO(6)$ symmetry down to $SO(5)$, and we assume that the coefficients $\lambda$ are order 1.  Additional global symmetries can be explicitly broken by choosing a more general $\lambda_{ij}$.  To achieve stable DM, it is important to assume that the potential at least preserves an accidental global $SO(2)$ flavor symmetry.

The equations of motion for the $S$ fields set all meson VEVs to be zero, which contradict the quantum modified constraint
\begin{equation}
M_0M_0+\sum_{i}M_iM_i=\Lambda^2.
\label{eq:itiy-constraint}
\end{equation}
Thus SUSY is broken.  Analogous to \Sec{subsec:32explicit}, we assume that the vacuum preserves the maximal $SO(5)$ global symmetry, such that of the meson, only $M_0$ obtains a VEV $\langle M_0 \rangle = \Lambda$.  Therefore, the fields $M'=M_i$ and $S' = S_i$ for $i=1,...5$ are stable DM candidates since they are charged under the (unbroken) $SO(5)$ global symmetry. 

Note that $M_0$ carries no $R$-charge, so unlike the 6-1 model in \Sec{sec:32}, there is no guarantee that the $R$-symmetry is broken.  Below, we will make the assumption that $S_0$ does get a VEV, in order to have the desired phenomenology.

\subsection{Spectrum and annihilation}

One can solve the constraint in \Eq{eq:itiy-constraint} by writing $M_0=(\Lambda^2-M^{\prime2})^{1/2}$. The resulting effective superpotential is 
\beqs
W_{eff}&=&\lambda_0 S_0(\Lambda^2-M^{\prime2})^{1/2}+\lambda_1 S^\prime M^\prime,  \nonumber \\
& =& \tilde{\lambda}_0 \Lambda^2 S_0 - \tilde{\lambda}_0^\prime S_0M^{\prime2}/2+\tilde{\lambda}_1\Lambda S^\prime M^\prime+{\cal{O}}(M^{\prime4}),
\eeqs
where again $S^\prime=S_i, M^\prime=M_i$ for $i=1,...5$. In the second line, we have canonically normalized the composite fields and the coefficients $\tilde{\lambda}$ absorb unknown factors from the (non-canonical) K\"{a}hler potential.  In this form, is clear that the ITIY model is a type of O'Raifertaigh model. By NDA, we have 
\beq
\tilde{\lambda}_0 \sim g, \quad \tilde{\lambda}_0^\prime \sim g^{-1}, \quad \tilde{\lambda}_1 \sim 1,
\eeq
which is obtained after we take the $\lambda_i$ to be order one and $g \sim (4\pi)^{-1}$.

After SUSY breaking, there is one linear combination of the singlets $S$ whose VEV is undetermined assuming the minimal quadratic K$\ddot{\rm{a}}$hler potential for $S$. This is the tree-level flat direction of any O'Raifertaigh model.  Previous analysis~\cite{extended itiy} has shown there is a minimum at $\langle S \rangle =0$ by calculating the one-loop contributions from the light particles.  Such an analysis is valid in the weakly-coupled regime, but for strongly coupled DM this conclusion may be modified.  For example, when $\langle S\rangle \sim \Lambda$, the one-loop analysis breaks down as all loops contribute at the same order. Thus one cannot exclude the possibility that there is a (local) $R$ symmetry breaking vacuum in the strongly-coupled region $\langle S\rangle \sim \Lambda$, as assumed in Ref.~\cite{extended itiy, Ibe:2007wp}.  We assume that such a vacuum exists with $\langle S_0 \rangle \equiv v \sim \Lambda$ while $\langle S^\prime \rangle = 0$, such that the model still preserves the original $SO(5)$ global symmetry.  As in \Sec{sec:32}, the unstable field $S_0$ will then decay to $R$-axions.

Expanding around such a vacuum, the effective superpotential is
\beq
W_{eff} =  \tilde{\lambda}_0 \Lambda^2 \tilde{S_0}- \tilde{\lambda}_0^\prime (v+\tilde{S_0}) M^{\prime2}/2+\tilde{\lambda}_1\Lambda S^\prime M^\prime,
\eeq
where $\tilde{S_0} = S_0 -v$. We see that $S^\prime$ and $M^\prime$ get a vector-like mass of order $\Lambda$. The field $\tilde{S_0}$ will receive masses from higher order K\"{a}hler corrections such as $|S_0|^4$. Assuming that  $\tilde{S_0}$ is lighter than the lightest component of $M^\prime$, the annihilation channel could be $M^\prime M^\prime \to \tilde{S_0}\tilde{S_0}$. The interaction terms generated from the superpotential such as $V \sim \tilde{\lambda}_0^{\prime 2} |\tilde{S_0}|^2|M^\prime|^2 + \cdots$ lead to an annihilation cross section of the order $\tilde{\lambda}_0^{\prime 4}/\Lambda^2 \sim (4 \pi)^4/(100$~TeV)$^2$ which has the right parametrics to yield the correct dark matter relic abundance.

Thus, as long as we assume that a strongly coupled ITIY-type model dynamically breaks an $R$ symmetry, then it has the right phenomenology to produce DM in the DSB sector.

\subsection{Dark matter decays}
\label{subsec:DMdecayITIY}

As in \Sec{sec:32decays}, DM could decay on cosmological timescales via flavor-violating operators suppressed by a high scale.  The leading flavor-violating operators are the K\"{a}hler potential terms
\beq
K= \frac{1}{M_*^2} \eta^{ab}_{cd}Q_a^\dagger Q^c Q_b^\dagger Q^d,
\eeq
with arbitrary flavor structure for $\eta^{ab}_{cd}$. Below the confining scale $\Lambda$, this operator is mapped onto 
\beq
K=\frac{\Lambda^2}{M_*^2}c^{ab}M^{\dagger}_aM_b, 
\eeq
with coefficients $c^{ab}$. For $\eta \approx {\cal{O}}(1)$, $c \approx {\cal{O}}(1)$ by NDA.  This will induce mixings between different components of the mesons. 

Consider one example $K=(\Lambda^2/M_*^2) M^{\dagger}_0M^\prime$. By a field redefinition of $M_0$ that diagonalizes the kinetic terms, one would have in the superpotential an $SO(5)$ symmetry breaking operator
\begin{equation}
\delta W= \frac{\Lambda^3}{M_*^2}  S_0 M^\prime.
 \end{equation}
This forces the fields charged under the global symmetry to obtain tiny VEVs 
\beq
\langle M^\prime, S^\prime \rangle = c^\prime  \frac{\Lambda^3}{M_*^2},
\eeq
where the coefficient $c^\prime \approx {\cal{O}}(1/4\pi)$ by NDA. Even though $M^\prime$ is not charged under the $R$ symmetry, its scalar mixes with the $R$-charge two singlet $S^\prime$ through the mass terms in the potential as $\Lambda^2 M^\prime S^\prime$. Thus the $M^\prime$ scalar can decay into two $R$-axions while its fermionic partner can decay into an $R$-axion and a gravitino.  The estimate of the lifetime is similar to that of the DM in 6-1 model.  Another possibility for the scalar DM is that $M^\prime \to \tilde{S_0} \tilde{S_0}$ if it is kinematically allowed, $\mathrm{mass}(M^\prime) > 2 \, \mathrm{mass}(\tilde{S_0})$. Here the $\tilde{S_0}$ scalars in the final state will subsequently decay to $R$-axions.

\section{Gauge mediation}
\label{sec:messengers}

Having shown that viable DM can exist in the DSB sector, we now briefly discuss how to incorporate such models into realistic gauge mediation scenarios.  The goal is to introduce messenger fields with SM gauge charges which directly feel the SUSY breaking of the DSB sector.  Yukawa couplings of messengers to the composite fields in the DSB sector are typically highly suppressed since such couplings are high dimension operators from the point of view of the microscopic DSB fields.  Thus, we are led to introduce messengers which are directly charged under the hidden gauge group. 

For instance, in the 6-1 model from \Sec{sec:32}, one could add messengers charged under the strong $SU(6)$ group~\cite{Ibe:2007wp, direct}.\footnote{One could also try adding a pair of messengers charged under the gauged $U(1)$ in the 6-1 model.  However, since $U(1)$ is weakly coupled at the DSB scale, the mass splittings of such messengers would be one-loop suppressed from $\Lambda$.  Consequently the MSSM soft masses are two-loop suppressed from $\Lambda$, in essence yielding a two-step weakly coupled gauge mediation. To have acceptable soft masses, $\Lambda$ has to be $(10^4 - 10^5)$ TeV, which is unacceptably large from the point of the view of achieving DM in the DSB sector.}  Specifically, we add messengers $P$, $\tilde{P}$ which transform as a fundamental and anti-fundamental under $SU(6)$ with a superpotential mass term $m_P P\tilde{P}$.  For a SUSY breaking scale $\sqrt{F} \sim 10 -100~\mathrm{TeV}$, $m_P$ must taken to be close to the strong-coupling scale $\Lambda$.   At the scale $\Lambda$, one expects the $SU(6)$ strong dynamics to generate K\"{a}hler operators of the form $|P|^2|\Phi|^2$ where $\Phi$ are the composite moduli.  Such operators lead to SUSY breaking mass splittings inside the messenger sector of order $\Delta m \approx {\cal{O}}(\Lambda)$. 

These messengers transmit SUSY breaking from the DSB sector to the SSM because they are charged under the SM gauge group, e.g.\ $P+\tilde{P}$ transform as $5+\bar{5}$ under the GUT group $SU(5)_{\rm SM}$.   With this messenger content, $SU(6)$ is still asymptotically free, so the desired DSB dynamics persists.  The low energy MSSM soft masses can be parametrized by the current-current correlators of the hidden sector as in general gauge mediation~\cite{general}.  By NDA, 
\begin{eqnarray}
m_{1/2}& \sim &\frac{g^2}{4\pi}\frac{F}{\Lambda} \sim \frac{g^2}{16\pi^2}\Lambda, \nonumber \\
m_0^2 &\sim& C_2(r)m_{1/2}^2 \sim C_2(r) \left(\frac{g^2}{16\pi^2} \right)^2 \Lambda^2,
\end{eqnarray}
where we used $F \sim \Lambda^2/4\pi$ and $C_2(r)$ is the quadratic Casimir invariant of the representation $r$ of the scalar.  Here the factors of $4\pi$ arise from NDA and are coincident with the loop factor in the weakly-coupled case.  $\Lambda \sim$ (10 -- 100) TeV gives the MSSM gaugino mass around (100 GeV -- 1 TeV), as required.

Note that adding these six sets of $5+\bar{5}$ messengers will ruin perturbative GUT unification.   Of course, preservation of perturbative grand unification is not a necessary requirement, and in fact, DSB sectors with strongly coupled messengers tend not to have this property.  However, in the context of the decaying DM scenarios envisioned in \Sec{subsec:generalDMdetection}, perturbative unification is a desirable feature.  The reason is that to get DM with cosmological lifetimes, the higher dimensional symmetry breaking operators must be suppressed by the GUT scale, and without perturbative unification, the motivation for such operators is lost.  Possible ways of preserving perturbative gauge unification despite the large number of messenger fields have been discussed in Ref.~\cite{messenger}.  

One marginal possibility in the class of dynamical superpotential models is to consider the 5-2 construction, i.e. the DSB model $SU(n) \times SU(2) \times U(1)$ ($n$ odd) with $n = 5$.  Such a model can also yield strongly coupled DM, but need only have five $5+\bar{5}$ messengers to achieve gauge mediation.  With $\Lambda \sim (10 - 100)~\mathrm{TeV}$, unification is marginally preserved~\cite{jones}.  We emphasize, though, that having DM decay is not necessary from the point of view of having a viable DM scenario, but is simply an interesting possibility in light of indirect DM detection.

Turning to models with a quantum moduli space, note that adding $5+\bar{5}$ messengers does not work for the simplest ITIY model in \Sec{sec:itiy}, as it ruins the asymptotic freedom of the gauge group $SU(2)$.  To add strongly coupled messengers, we must consider an ITIY-type model with a larger gauge group, for example an ITIY model based on a $Sp(2N)$ gauge group with $2N+2$ fundamentals $Q$ and $N(2N-1)$ gauge singlets~\cite{extended itiy}. 
The effective superpotential is a generalization of the one for the $SU(2)$ model (see Eq.~(35) in Ref.~\cite{extended itiy}) and can be analyzed in the same way.  For $N>1$, asymptotic freedom is preserved after adding messengers $P(\tilde{P})$ in the (anti)fundamental representation of $Sp(2N)$~\cite{Ibe:2007wp}.  $N=2$ is the unique choice for which perturbative unificiation is preserved after adding the messengers. 

Finally, messenger fields may themselves influence cosmology.  The simplest messenger sectors preserve a messenger parity, so the lightest component is stable and may provide an additional DM component.  Since these messengers are charged under the DSB gauge group, they experience strong annihilation cross sections, and as long as the lightest messenger is heavier than some unstable fields in the DSB sector, their relic abundance will be comparable to the other stable DSB fields.  However, since the messengers are charged under the SM gauge group, there will always be a colored component, and there are strong constraints on colored particles with lifetimes $\tau \ge 10^{17}$ sec as they form exotic atoms~\cite{Hemmick:1989ns}. 
Breaking the messenger parity with renormalizable interactions between the messenger and observable sectors could reintroduce the flavor problem.  An option which preserves messenger parity is to make the triplet inside $5+\bar{5}$ decay through a dimension-five operator suppressed by the Planck scale, e.g., $\int d^2\theta P^2 \bar{5}^2$, where $\bar{5}$ refers to a SM multiplet~\cite{Arvanitaki:2008hq}.

\section{Conclusions}
\label{sec:conclude}

Most studies of DM focus on the weakly-interacting DM paradigm, where DM has mass around the electroweak scale $m_{\rm DM} \sim {\cal{O}}(100 \, \rm{GeV} - 1\, \rm{TeV})$ and couplings of weak interaction strength $\kappa \sim {\cal{O}}(1)$, leading to an annihilation cross section that yields the desired thermal relic density.  This paradigm, however, is difficult to incorporate into another theoretically attractive paradigm, GMSB, which solves the flavor problem of the SSM.  In standard GMSB, the LSP is the super-light gravitino, which cannot play the role of cold DM in the universe.

In this paper, we have explored an alternative mechanism via which GMSB could still yield thermal relic dark matter, by considering DM in the DSB sector.  Since the relic density of DM is controlled by the DM annihilation cross section, and since this cross section depends mainly on the ratio $\kappa^2/m_{\rm DM}$, it is possible to arrange $m_{\rm DM}$ to be related to a heavier scale than the weak scale provided the DM couplings are correspondingly increased. Unitarity of the annihilation cross section then bounds the DM mass to be ${\cal{O}}(10-100\, \rm{TeV})$ and couplings to be of order a few to $4\pi$.  A DSB sector with the SUSY breaking scale $\sqrt{F} \sim {\cal{O}}(10-100\, \rm{TeV})$ naturally realizes this situation, and heavy DM could be a composite state of SUSY breaking strong dynamics.  

We have found that DM can indeed arise from two of the most popular DSB scenarios:  chiral models with a dynamically generated superpotential and vector-like models with a quantum moduli space.  Both of these setups allow the DSB sector to have a large global symmetry that is preserved by SUSY breaking, and the lightest state(s) charged under this global symmetry are stable DM candidates.  This setup also provides an additional probe of the dark sector: the $R$-axion.  DSB models often preserve an approximate $R$ symmetry which is broken when SUSY is broken.  The resulting $R$-axion is naturally light, 1 MeV - 10 GeV, with an irreducible contribution to its mass from supergravity effects.  We have also argued that these DSB sectors can incorporate strongly-coupled messenger fields as necessary to have a complete model of GMSB.

These models are an explicit realization of the scenario envisioned in Ref.~\cite{Mardon:2009gw}, where decay of heavy composite DM might be responsible for various astrophysical anomalies.  Indeed, the symmetries that stabilize DM need not be respected by high scale physics at the unification or the Planck scale.  The compositeness of DM then forces the leading symmetry-violating operators to be high-dimensional, and consequently DM decays on cosmological time scales.  In particular, we have found explicit DSB models where the leading symmetry violating terms are dimension-six operators.  When suppressed by the Planck scale, such operators yield a DM lifetime of $10^{26}$ sec for $m_{\rm DM} \sim$ 10 TeV, a range that can be probed by current and future DM indirect detection experiments.  Since the leading DM decay is to $R$-axions, and since by kinematics the $R$-axion decays dominantly to SM leptons and pions without producing many hadrons, such a scenario can explain the lepton-rich astrophysical anomalies.

The possibility of DM in the DSB sector presents a new direction for SUSY model building.  While the bulk of previous DSB studies have focused on simply establishing SUSY breaking---the zeroth-order requirement for a DSB sector---it is worthwhile to develop tools to analyze the symmetries and spectra of the SUSY breaking vacuum, especially for strongly-coupled models.  We have shown that explicit DSB model have residual global symmetries and viable cold DM candidates, and we expect this to be true for a wide variety of DSB scenarios.  In our models, besides the SSM spectrum which encodes limited DSB information, light states such as $R$-axions provide additional handles on hidden sector dynamics which could be tested in near future astrophysics and collider experiments \cite{Mardon:2009gw,Nomura:2008ru,raxion_pheno}.

\section{Acknowledgement} 
We thank David Krohn for collaboration at early stages of this project. L.-T.W. is supported by the National Science Foundation under grant PHY-0756966 and the Department of Energy (D.O.E.) under Outstanding Junior Investigator award DE-FG02-90ER40542.  J.T. is supported by the D.O.E. under cooperative research agreement DE-FG0205ER41360.

\end{document}